# High Pressure Effects on Superconductivity in the $\beta$-pyrochlore Oxides AOs$_2$O$_6$ (A=K, Rb, Cs)

Takaki MURAMATSU, Shigeki YONEZAWA, Yuji MURAOKA and Zenji HIROI

*Institute for Solid States Physics, University of Tokyo, Kashiwa, Chiba 277-8581*

*E-mail address: muramatu@issp.u-tokyo.ac.jp

Superconductivity in transition metal oxides (TMOs) has been one of exciting fields in solid state physics due to its unconventional properties. Many superconductors in TMOs crystallize in perovskite related structures. Recently, an exceptional TMO superconductor Cd$_2$Re$_2$O$_7$ was found with $T_c$ = 1.0 K, which crystallizes in the pyrochlore structure.[1] The pyrochlore structure contains a corner-sharing tetrahedral network made of transition metal cations called the pyrochlore lattice, and is known as one of spin frustration systems in the case that the transition metal cation has a localized moment with antiferromagnetic interactions between nearest neighbors. Very recently, three related compounds AOs$_2$O$_6$ (A = K, Rb, Cs) named the $\beta$–pyrochlore oxides were found,[2-4] which contain the pyrochlore lattice made of Os atoms. They show superconductivity at $T_c$ = 9.6 K, 6.3 K and 3.3 K, respectively. Particularly in KOs$_2$O$_6$ the $T_c$ is almost one order higher than in Cd$_2$Re$_2$O$_7$. The mechanism of the superconductivity has been extensively studied and was suggested to be unconventional. For example, a remarkably large upper critical field $H_{c2}$ of 38 T exceeding Pauli's limit was reported for KOs$_2$O$_6$.[5] Moreover, the μSR and NMR experiments suggested an anisotropic order parameter.[6,7] On the other hand, there are a few reports which suggest conventional BCS superconductivity for RbOs$_2$O$_6$.[8,9]

It is plausible to assume that the systematic change of $T_c$ in AOs$_2$O$_6$ is due to the size effect of alkaline metal ions. The lattice constant is almost proportional to the ionic radius of A ions, and the $T_c$ increases with decreasing the lattice constant from Cs to K under a positive chemical pressure.[4] This is in contrast to the case of conventional BCS superconductivity in a single-band model, where the $T_c$ decreases under a positive pressure, because the density of state (DOS) decreases with decreasing the lattice volume, as typically observed in alkali metal doped C$_{60}$ superconductors A$_3$C$_{60}$.[10] This contrast may indicate the uniqueness of the superconductivity in AOs$_2$O$_6$. Thus, it is interesting to examine physical pressure effects on $T_c$. In this note, we report diamagnetic measurements under pressures up to 1.2 GPa using a piston-cylinder type pressure cell on the three compounds of AOs$_2$O$_6$. In the $\alpha$-pyrochlore oxide superconductor Cd$_2$Re$_2$O$_7$ the high pressure study revealed that $T_c$ increases from 1.0 K to 3.0 K at 2 GPa and then decreases to vanish above 3 GPa.[11,12] In the case of RbOs$_2$O$_6$, Khasanov *et al.* reported a high pressure experiment in which the $T_c$ increases monotonously up to 1.0 GPa with the initial slope of 0.90 K/GPa.[8]

Polycrystalline samples were prepared as reported previously.[2-4] They were nearly single-phase, but with a small amount of OsO$_2$ or Os metal. Magnetic susceptibility was measured in a Quantum Design MPMS. Quasi-hydrostatic pressures up to 1.2 GPa were produced by a piston-cylinder type pressure cell which is made of hardened CuBe alloy. Daphne oil 7373 was used as a pressure transmitting medium. The actual pressure was determined in each experiment by measuring the superconducting transition temperature of Pb or Sn which was put into the pressure medium together with a sample and placed about 2 cm away from the sample.

A typical temperature dependence of magnetic susceptibility for KOs$_2$O$_6$ is shown in Fig.1. In a field of 10 Oe after zero-field cooling, a large diamagnetic signal due to the shielding effect was observed at 4 K in the whole pressure range. On heating, it disappears rapidly above 9 K. We defined $T_c$ in two ways. One is the onset temperature where the magnetic susceptibility starts to deviate from a background signal at high temperature. This temperature corresponds to the $T_c$ determined by specific heat measurements or resistivity measurements on bulk samples at ambient pressure in a zero magnetic field. The other characteristic temperature $T_c$' is determined to be the temperature where the diamagnetic signal becomes 5% of the maximum shielding signal.

The pressure dependence of $T_c$ as well as $T_c$' is plotted in Fig.2. In RbOs$_2$O$_6$ and CsOs$_2$O$_6$, the $T_c$ increases almost linearly with increasing pressure in the whole pressure range examined. The slope for RbOs$_2$O$_6$ is 0.78 K/GPa, which is slightly smaller than the value reported by Khasanov.[8] In remarkable contrast, the $T_c$ of KOs$_2$O$_6$ exhibits a downturn with a maximum value of 10.0 K at 0.56 GPa, and then decreases to 9.5 K at 1.20 GPa. In order to compare the three systems, the pressure dependence of $T_c(P)$ normalized by the $T_c$ at ambient pressure (AP) is shown in Fig. 2(b). The initial slope of $\{T_c(P)/T_c(AP)\}/P$ is 0.20, 0.14 and 0.13 GPa$^{-1}$ for A = Cs, Rb and K, respectively; the largest for CsOs$_2$O$_6$ with the largest unit cell volume and the lowest $T_c$ at AP. The initial increase of $T_c$ by applying physical pressure commonly observed for the three compounds is consistent with the trend by chemical pressure. This means that the lattice volume is the key parameter to determine the $T_c$ of AOs$_2$O$_6$. The origin of the saturation and the following downturn in $T_c$ for KOs$_2$O$_6$ is not clear at the moment but may indicate that a fluctuation relevant to the superconductivity is enhanced at certain pressure and suppressed of higher pressure. Further experiments at higher pressures are necessary to discuss this issure in more detail and are in progress.

In conclusion, we measured the pressure dependence of magnetization up to 1.2 GPa in order to deduce the pressure effect of $T_c$ in the three *β*-pyrochlore oxides. It is found that the initial pressure dependence of $T_c$ is positive for all the compounds. Only KOs$_2$O$_6$ exhibits a saturation in $T_c$ at 0.56 GPa and the downturn at higher pressure.

This work is supported by a Grant-in-Aid for Scientific Research (14750549) given by the Ministry of Education, Culture, Sports, Science and Technology. T. M. thanks for a financial support by a Grant-in-Aid for Creative Scientific Research (13NP0201)

Figure caption

Fig. 1
Temperature dependence of magnetic susceptibility of $KOs_2O_6$ measured at ambient pressure (AP) and high pressures of $P$ = 0.56 and 1.20 GPa. The measurements were carried out on heating after zero-field cooling in a magnetic field of 10 Oe.

Fig. 2.
Pressure dependence of $T_c$ (a) and nomalized one by the value at AP (b) for $AOs_2O_6$ (A=K, Rb, Cs). The solid and open marks represent the onset $T_c$ and $T_c'$.

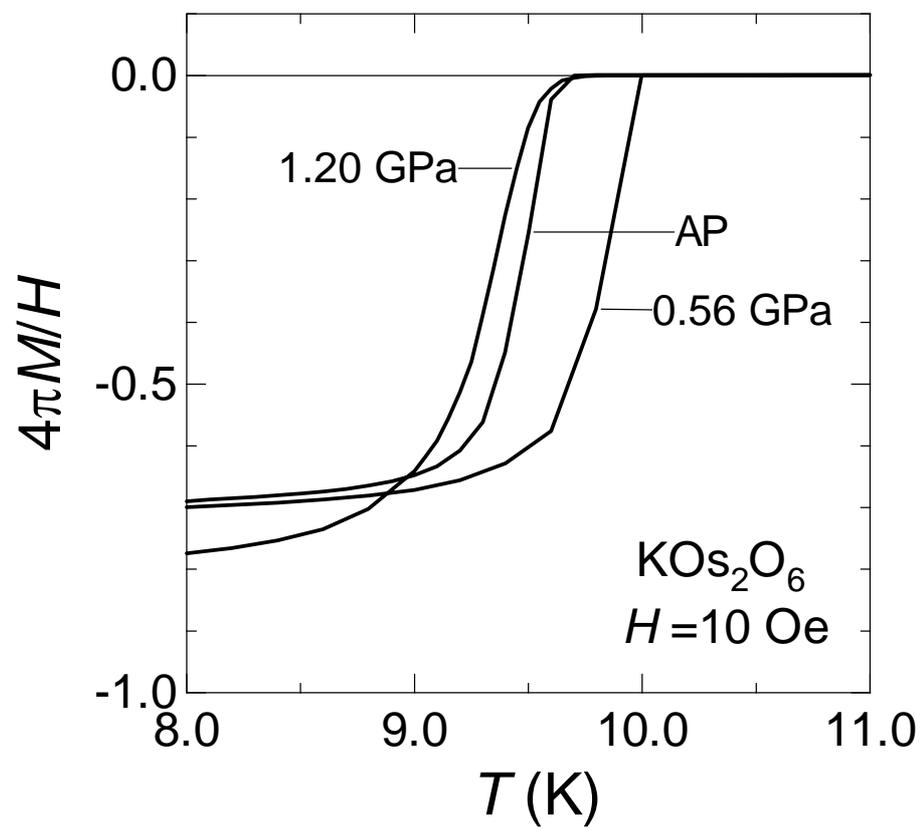

T. Muramatsu *et al*.

Fig. 2

(a)

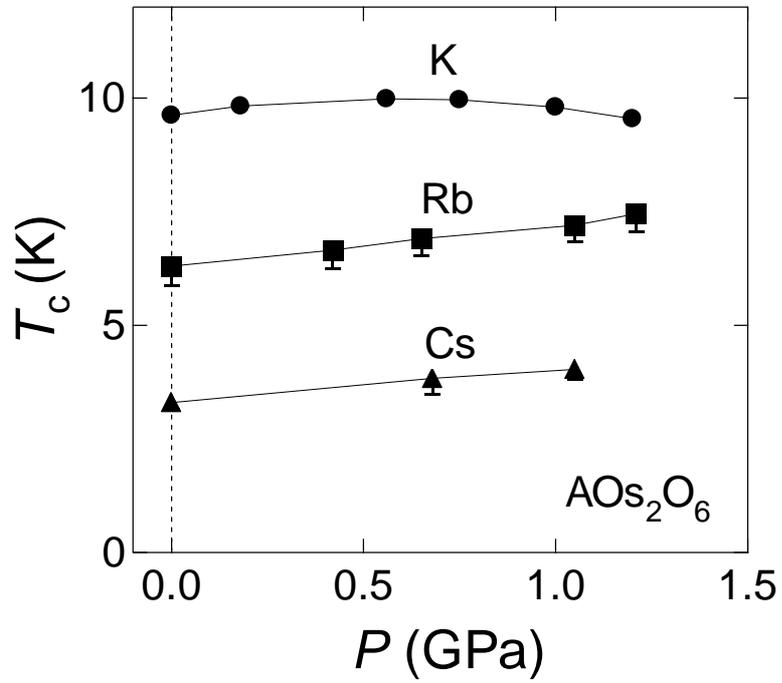

(b)

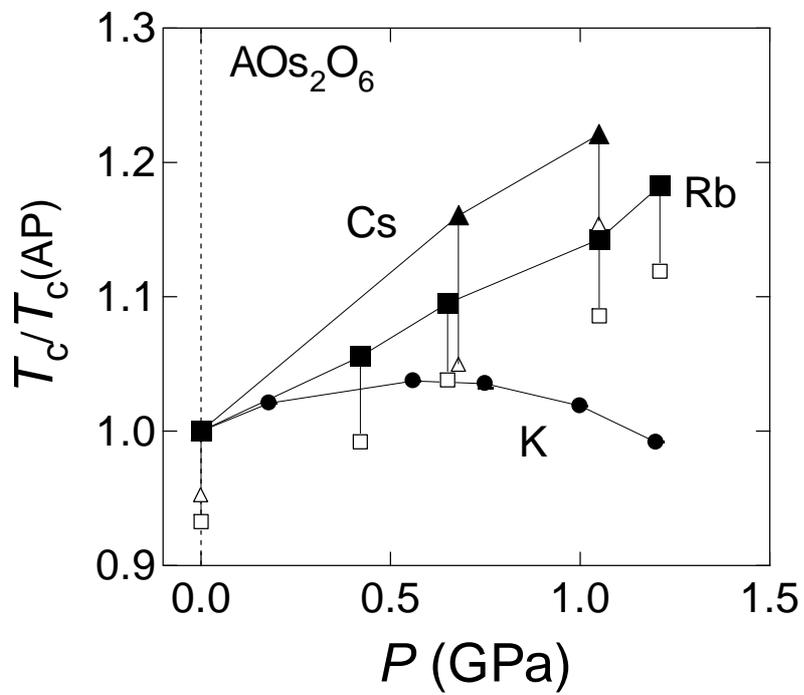